# Metal organic interfaces at the nanoscale

Cedric Troadec[1], Deng Jie[1], Linda Kunardi[1], Sean O' Shea[1], N Chandrasekhar[1,2]

[1]Institute of Material Research and Engineering (IMRE), 3 Research Link, Singapore 117602, Singapore

[2]Department of Physics, National University of Singapore, 2 Science Drive 3, Singapore 117542, Singapore

**Abstract**

In this work, we present an investigation of the Ag-PPP (polyparaphenylene) interface using ballistic electron emission microscopy. Our work is the first successful application of the BEEM technique to metal-organic interfaces. We observe nanometer scale injection inhomogeneities. They have an electronic origin, since we find corresponding Schottky barrier variations. We also determine the transmission function of Ag-PPP interface and find that it agrees qualitatively with the theoretical calculations for a metal-phenyl ring interface. We conclude that charge transport across inhomogeneous barriers needs to be considered for understanding electronic transport across metal-organic interfaces and organic device characteristics.

**Introduction**

Metal-organic (MO) interfaces have traditionally been investigated by I-V, C-V and UV spectroscopy, all of which average over macroscopic areas [1]. In contrast, prototype devices incorporating molecules as active components are sub-micron [2-4]. Recent work [4] has shown that nanoscale conductance inhomogeneities exist at all MO interfaces.

This has been ascribed to nanoasperities, since conductance increase has been observed over lengths scales of a few tens of nanometers in metal-molecule-metal structures. The physical origin of these nanoasperities or filaments remains obscure. Filament growth and dissolution has been identified as being responsible for the switching behavior in other systems as well [5,6]. Therefore filament formation is not exclusive to organic devices. Devices with dissimilar materials as electrodes have been essential for obtaining reversible switching. Switching can be due to electrochemical reaction [5], thermal migration and/or electromigration [6]. Lau et al. [4] report the observation of single switching center, and suggest a runaway process of filament growth driven by increasing current density and/or electric field. Memory effects observed in inorganic semiconductors [7] have been invoked to explain the behavior of some organic devices [8,9]. Organic device configurations that have been investigated to date are either Langmuir Blodgett (LB) films [4] or self-assembled monolayers (SAM) [2,3]. At the present time, it is unclear whether the inhomogeneities originate from microstructural perturbations such as asperities at the interfaces with the contacting electrodes, or whether they are an inherent electronic property of MO interfaces. Furthermore, SAM and LB films are not quite rigid, and despite the implementation of precautionary measures, it is uncertain whether the integrity of the organic is maintained after deposition of the metal film [10]. For instance, in metal-inorganic semiconductor (MIS) interfaces, unless the semiconductor surface is prepared with care and the metal is chosen so that it is lattice matched, the metal film is polycrystalline, causing significant variations in the electronic transparency of the interface [11,12].

In an extensive study of different molecules and metals, Whitesides and co-workers [13] saw no evidence that simple aromatic groups differ fundamentally in their behavior from aliphatic groups. They observed that aliphatic and aromatic systems with similar thickness have similar values of breakdown voltages (BDV). This raises a question of whether the breakdown mechanism is identical for all organics, and independent of the constituent groups [13]. Answering this question requires knowledge of the potential profile across the metal-organic-metal structure, as we show below.

Transport through metal-molecule-metal junctions is expected to depend critically on the properties of the contacts, which in turn depend on the details of the chemisorption or physisorption at the MO interfaces. The influence of the contacts can be described by the electrostatic potential profile across the junction. Datta [14] has concluded that to first approximation the change in potential is not linear, but displays sharp drops at the electrode-molecule interfaces. Equal potential drops at both contacts were necessary to account for symmetric *I-V* characteristics observed in STM conductance studies of SAM. A recent self-consistent solution of the Schrodinger and Poisson equations for a metal-molecule-metal junction by Mujica et al. [15] supports Datta's original conclusion that the potential drops at the electrode-molecule interfaces, with an approximately linear variation along the length of the molecule. In contrast, a density functional theory (DFT) calculation by Lang and Avouris [16] of the electrostatics for a metal-cumulene-metal junction shows no steep changes in the potential profile, but a large fraction of the total potential is reduced just *inside* the metallic contacts. In either case, the upshot is that a large fraction of the applied potential is dropped at (or within) the contacts, meaning that the contacts represent a significant bottleneck to current flow. These findings are similar to the Schottky barrier (SB) for MIS interfaces.

Clearly, the height and shape of injection barriers remain crucial for a better understanding of the transport through molecules or organics sandwiched between metal electrodes. Furthermore, Whitesides et al. [13] found that the highest value of BDV observed for both aliphatic and aromatic SAM is indistinguishable from that for thick (1 mm) films of polyethylene. Mechanical stress can cause significant changes in the properties of a MO interface [13,17]. However, mechanical failure of the polyethylene is certainly not involved in the breakdown of that material. Therefore, Whitesides et al. [13] conclude that organics share a common mechanism for breakdown, involving electronic current. These observations highlight the importance of an electronic study of the MO interface with sub-micron resolution, which we present in this work, for the Ag-PPP interface using BEEM. Our work is the first successful application of the BEEM technique to MO interfaces.

**Experiment**

In this work, we probe charge transport across the Ag-PPP interface with nanometer resolution, using ballistic electron emission microscopy (BEEM). The choice of materials is driven by relevance, ease of availability and experimentation. Our choice of materials overcomes the concerns listed by Bao et al. [10], since we do not work with SAM. It is well established [18] that phenylene oligomers can be sublimed onto cold substrates to yield films, whose properties are almost identical to those of the powder. A device configuration and schematic for BEEM, is shown in Fig. 1. An organic semiconductor is overlaid with a thin metal film (typically < 10 nm, termed the base), with an ohmic contact on the opposite side (termed the collector). The top metal film is grounded, and carriers are injected into it using a scanning tunneling microscope (STM) tip. These carriers are injected at energies sufficiently high above the metal's Fermi energy, so that they propagate ballistically before impinging on the interface. There is spreading of carriers in the metal film due to mutual Coulomb repulsion as well as some scattering by imperfections. When the energy of the carriers exceeds the Schottky/injection barrier, they propagate into the semiconductor and can be collected at the contact on the bottom. Typically the tunneling current is attenuated by a factor of 1000, so that collector currents are in the picoampere range. Spectroscopy and imaging can be done on this structure, by monitoring the collector current as a function of STM tip bias voltage at a fixed location, or as a function of tip position at a fixed STM tip bias. One of the fundamental advantages of BEEM is the ability to investigate transport properties of hot electrons with high lateral resolution, possibly at the nanoscale.

Phenylene oligomers, in the form of yellowish flakes were evaporated onto a pre-deposited gold film on a glass substrate, which serves as the collector. Choice of the base depends on the injection barrier that is to be measured. We have chosen Ag, since PPP is a hole transport material, and the Schottky-Mott rule yields a barrier of 0.9 V for the Ag-PPP interface. However, this ignores band bending, and the possibility of dipoles at the MO interfaces, which change its magnitude [19]. Evidence for the existence of interfacial dipoles comes from various experiments, including UPS and Kelvin probe. The Ag film is nominally 10 nm thick. The experiments were done at 77 K in a home-assembled STM system. The current noise of the setup is typically 1 pA. Ag has been shown to yield "injection limited" contacts for hole injection into the polyparaphenylene/vinylene family of organics [19]. It is important to ensure that the Ag film is reasonably flat, since the BEEM actually grounds the area of the metal investigated by the tip. Unless this requirement is met, attempts to tunnel into patches of the metal film, which are poorly connected, can lead to tip crashes.

**Results**

Prior to BEEM, SEM, TEM and FTIR studies of PPP films were done to determine the structure. The SEM images are unremarkable and indicate a reasonably flat film, with no pinholes or protrusions. TEM (see Fig. 2) indicates the presence of domains 50 nm wide and 100 to 300 nm long in an amorphous matrix. They seem to have no preferred orientation and have irregular boundaries. After deposition of the Ag film on top, it is clearly impossible to discern whether or not a domain or domain boundary is underneath an STM scan area. FTIR (not shown) indicates that the number of phenyl rings in the film is between 4 and 7.

The STM tip is used to inject holes, making this a ballistic hole emission spectrum. Figure 3 shows a raw current-voltage (I-V) spectroscopy over a 0 to 2 V range. Over two hundred individual I-V's acquired over different locations within a 25 nm square are averaged to obtain the black curve to reduce noise. Repeated acquisition of spectra at the same point were found to be detrimental to the sample, as evidenced by instability of the spectrum. Qualitatively, this curve is similar to BEEM spectra seen for MIS interfaces. The Schottky or injection barrier is usually taken to be the point where the collector current begins to deviate from zero. Early BEEM experiments on MO interfaces yielded low currents below 5 pA [20]. We believe that this is sample dependent, with spin-coated samples yielding almost undetectable currents.

Extraction of the SB from BEEM data, such as that shown in Fig. 3 requires modeling of the spectral shape. Bell and Kaiser [21] used a planar tunneling formalism for determining the shape. The solid green line is a Kaiser-Bell fit to the raw data, and has the form:

$$I_b = A (V-V_o)^n$$

(1)

where $I_b$ is the collector current, and $V_o$ is the injection barrier. We find that the value of $V_o$ ranges from 0.3 to 0.5. This should be contrasted with the injection barrier determined by the Schottky-Mott (SM) rule, which yields a value of 0.9 V. The exponent n varies from 2.76 to 3.13. The $V_o$ values, extracted in this manner, are shown as a histogram in Fig. 4. The substantial deviation of the barrier from the SM rule, and its distribution are noteworthy. We will discuss this below.

An STM image of the top Ag film, at 0.5 V and 1 nA is shown in Fig. 5(a). The I-V and dI/dV enables a choice of imaging conditions suitable to the particular interface. For instance, based on the spectroscopy data, it is possible to determine that bias voltage of 1 V should yield a measurable collector current. Plots of the collector current as a function of the STM tip position are images of electronic transparency of the interface. Such an image is shown in Fig. 5(b). The image clearly indicates non-uniform transparency of the interface over the region scanned by the STM. The bright spots indicate transparent regions. The size of such regions appears to be in the range of ten nanometers and above. BEEM studies of MIS interfaces often show a correlation between the STM, STM derivative and BEEM images. This is due to lateral variation of the surface density of states [11,12]. In this work, the electronic transparency of the interface and the surface morphology of the Ag film appear to be uncorrelated, a fact readily apparent from the images.

Figure 5(c) shows a profile of the BEEM current along the red line shown in the BEEM current image. Substantial variation in the BEEM current is readily apparent from one location to another. The BEEM current is a direct measurement of the conductance of the interface since it is a measure of the number of carriers being collected at the bottom electrode, at a fixed voltage which is applied to the STM tip. It is possible to estimate the electric field at the interface, assuming appropriate physical parameter values for PPP, using standard equations from semiconductor physics [21]. For a dielectric constant of 3, and a carrier concentration of $10^{13}/cm^3$ we obtain a field of $10^5$ V/m. We note that these fields are at least two orders of magnitude lower than the fields typically applied to organic devices. Several points are to be noted in comparing this work to that of Lau et al. [4]. We discuss these below.

**Discussion**

The voltages applied to the STM tip are comparable to the voltages applied by Lau et al. [4] to their LB organic devices. First, since the BEEM technique is based on STM, lateral resolution is much higher. For instance fluctuations in the BEEM current can be seen over 5 nm lengths scales. Secondly, the size scale of the inhomogeneities is of the same order. Thirdly, the order of magnitude change in conductance in our work, as measured by the BEEM current is indeed comparable to that reported by Lau et al. For one particular device, Lau et al. report a change of almost six orders of magnitude. Although we have not seen a corresponding change in the BEEM current, we believe that this is due to the differences in the experimental technique. We believe that the variations in the SB, as shown by the histogram in Fig. 4 can be used to reconcile these differences in the following manner.

Even in established Si and GaAs microelectronics, the properties of interfaces remain an active area of research. Recent work [22-24] suggests that a formation mechanism of the SB, which is strongly dependent on the local microstructure of the interface, explains experimental observations in a consistent manner. Schottky contacts on wide band-gap materials often exhibit significant device-to-device variations and/or ''non-ideal'' behavior in measured current-voltage (*I-V*) characteristics. This "physics" is "generic" to wide band-gap semiconductors, which organics are. Several studies [25-27] have demonstrated that small interfacial ''patches'' with reduced *local* Schottky-barrier height (SBH) of size comparable to or less than the semiconductor Debye length, cause a ''potential pinch-off'', i.e. current is funneled through the low barrier patch.

Since we have observed nanometer-scale patches by the BEEM technique, a discussion on how transport is influenced by them is important. Understanding the physical effects of these patches is essential for addressing issues relevant to organic electronics, such as switching, and breakdown. A model by Tung [22] considers an areal density of circular ''patches'' of low barrier height embedded in uniform background of barrier height $\Phi_o$. An individual patch is assumed to have a radius $r_o$ and a potential depth $\Delta$ relative to the background barrier height. Tung showed that a patch could (in certain limits) be treated approximately as an electrostatic *dipole* moment $p$ of strength $p = \varepsilon_s \Delta \pi r_o^2$, where $\varepsilon_s$ is the permittivity of the semiconductor. Although both high-barrier and low-barrier patches are present in the real system, the high-barrier patches can be neglected in calculation of excess current, since this excess current would fall off exponentially with barrier height, and the patches cover a relatively small fraction of the entire device area. Our BEEM current image, (Fig. 5(b)) shows that all these conditions are met by the Ag-PPP interface.

Injection across inhomogeneous interfaces can be quantitatively explained by assuming specific distribution of nanometer-scale interfacial patches of reduced SBH [24]. There is experimental evidence [28-30] that nanometer-sized lateral variations in SBH exist at many metal semiconductor interfaces. Olbrich et al. [28] have recently shown that potential pinch-off effect can be observed near intentionally introduced low-barrier height areas in Au/Co/GaAsP diodes, and that these low-barrier areas correlate well with the effective barrier height of the entire diode. In an analytical calculation, Tung et al. [24]

have shown that for a SB of 0.8 eV, nanometer scale patches that have a barrier reduced by 0.3 eV, result in currents that are three orders of magnitude larger, when compared to the ideal uniform barrier case. Such behavior is well known for MIS Schottky contacts, and is termed "non-ideality". In fact, it is common to speak of "ideality factors" in the MIS context.

The mechanism for inhomogeneous conduction across the interface is now evident. The BEEM current is "funneled" through regions that have a low SB. Our BEEM current profiles show variations of an order of magnitude as a function of position. Furthermore, the measured SB for the Ag-PPP interface has a variation of 0.2 eV, and ranges from 0.3 to 0.5 eV. Such a variation can easily cause nearly three orders of magnitude increase in current through the low SB patches, as shown by Tung et al. [24]. However, we do not observe three orders of magnitude variation in the BEEM current. The low mobility in the organic and the time scale of experimentation limit the observable BEEM current. Furthermore, in BEEM, the metal-organic-metal structure or device is not active. The base is grounded, and the collector is unbiased. If a drift field were to be applied across the organic, such large variations would be observable.

We also note that local variations in SBH can cause selective growth as reported by Schmuki and Erickson [31]. They report a technique that allows one to electrodeposit material patterns of arbitrary shape down to the submicrometer scale. An electrochemical metal deposition reaction can be initiated selectively at surface defects created in a *p*-type Si(100) substrate by Si focused ion beam bombardment. The key principle is that, for cathodic electrochemical polarization of *p*-type material in the dark, breakdown of the blocking SB at the semiconductor/electrolyte interface occurs at significantly lower voltages at implanted locations. This difference in the threshold voltages is exploited to achieve selective electrochemical deposition. These findings clearly indicate that the etching pretreatment exposes higher defect concentration and thus facilitates deposition. This suggests that metal nucleation and growth is either linked to a critical defect density or that for SB breakdown at sufficiently low potentials, a critical defect density is necessary. This leads to "filaments" of the same lengths scales as reported by Lau et al. [4].

Finally, we make a qualitative determination of the transmission function for the MO interface from our BEEM data. Temperature, disorder, applied field, and interface dipoles influence the transport of carriers across a MO interface [32-37]. Other variables are sample preparation technique, and time [38]. A successful model explains transport across such interfaces by considering hopping in the presence of a Coulomb potential [33-35]. The primary injection process from the Fermi level of the metal to the first layer of the organic is considered explicitly, while the subsequent diffusive random walk is treated as an Onsager-like process. Such a model predicts power law dependence of the current on the voltage. The primary injection then occurs in a small region of the interface comprising metal atoms and the first molecular layer of the organic. If this is indeed the case, it is possible to evaluate a transmission function for the Ag-PPP interface. We do this below.

The BEEM process is divided into three distinct steps. The first is the tunneling of the charges from the STM tip to the metal overlayer (the base). The second is the propagation through this metal layer, and last is the transmission across the interface [11,12]. Data for MIS interfaces has been analyzed with considerable success in this manner. Each of these processes is a function of the energy of the charges (electrons/holes). The product or convolution of these three functions yields the derivative of the BEEM spectrum, the dI/dV. The first step requires knowledge of spatial and energetic distribution of the electron current at the metal surface after tunneling. It depends on the density of states of the tip, the metal film, and the respective Fermi distributions, the temperature and the tip voltage. The functional dependence of this current on energy is well known [11]. The next step is attenuation/propagation in the metal film. This depends on the electron path length through the metal film and is energy dependent. Its functional form is an exponential decay [11]. The third step is the transmission across the interface, and depends on the energy and direction of the electrons. For MIS interfaces, free electron like behavior in the metal and the semiconductor, an energy independent and isotropic distribution of electrons in the metal film close to the interface is assumed. The transmission coefficient depends on the actual semiconductor band structure. In the case of BEEM for MO, the only parameter that changes is the transmission.

Retaining the functional forms of the tunneling current distribution, and the propagation through the metal film, taking their product, and dividing the dI/dV shown in Fig. 3 earlier, yields the transmission function (see Fig. 6). One interesting feature that is immediately apparent is that the MO transmission function has a curvature that is opposite to the curvature of a MIS transmission function. That is, the transmission function of the MO interface seems to scale with the square of the energy, in contrast to that of the MIS interface, which scales with the square root. The reasons for this behavior are unclear. However, this compares well with published literature, as we discuss below.

The atomic registry at the interface is unknown. PPP is composed of phenyl rings. Xue and Ratner [39] have studied the transmission across an Au-phenyl dithiol (PDT) and Au-bi-phenyl dithiol(BPD) structure. Most of the potential drop occurs at the Au-phenyl ring interface. Once charge is transferred from Au to the phenyl ring, further transport does not cause substantial potential drops, although it is accompanied by significant changes in the electron density on the molecules. In the case of the Ag –PPP interface, one can assume that the primary injection process is the transfer of charge from the Ag to the first phenyl ring of the PPP, in the absence of any evidence to the contrary. Therefore, comparing the normalized transmission function determined earlier with the calculation of Xue and Ratner [39] (Fig. 7), we find two common features: (a) both have a curvature that is concave upwards, i.e. they scale with a power of energy, which is greater than one, (b) both peak towards the HOMO levels of the organic. Given the crude approximations that have been made, the agreement between theory and experiment is noteworthy.

**Conclusion**

In conclusion, nanoscale inhomogeneities in SBH have been observed at metal-organic-metal junctions. Injection non-uniformity at MO interfaces is entirely electronic in origin and it is known to explain the behavior of MIS interfaces very well. Based on the literature reporting studies of electronically inhomogeneous MIS interfaces, knowledge of the precise atomic arrangement at MO interfaces would go a long way in aiding a detailed understanding of this problem. As in electromigration, the inhomogeneous SB may nucleate filamentary conduction driven by increasing current density and/or electric field. In addition, transmission function has been determined, which agrees reasonably well for an interface of a noble metal with a molecule composed of phenyl rings. Further investigations of other metals and organics are currently being addressed.

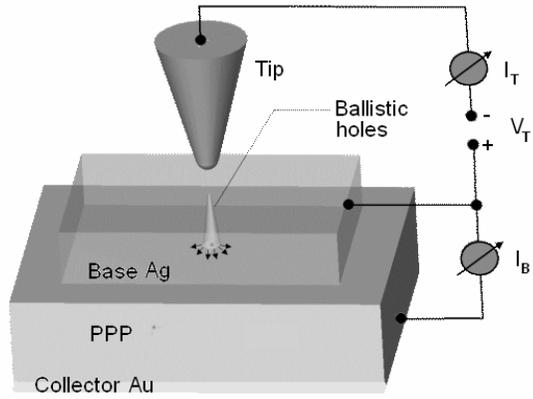

Figure 1. Schematic of the ballistic emission spectroscopy and imaging of an Ag-PPP buried interface. The PPP and Ag are evaporated onto a pre-deposited Au film on a glass substrate, held at 77 K to minimize interdiffusion.

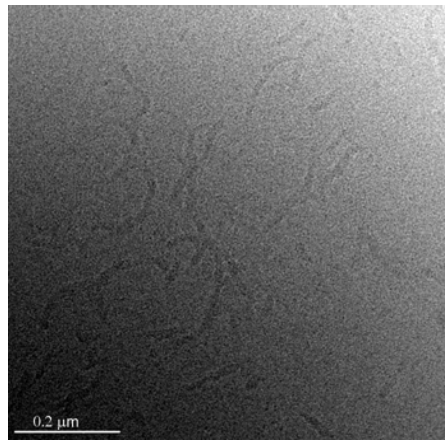

Figure 2. TEM image of a nominally 100 nm thick PPP film.

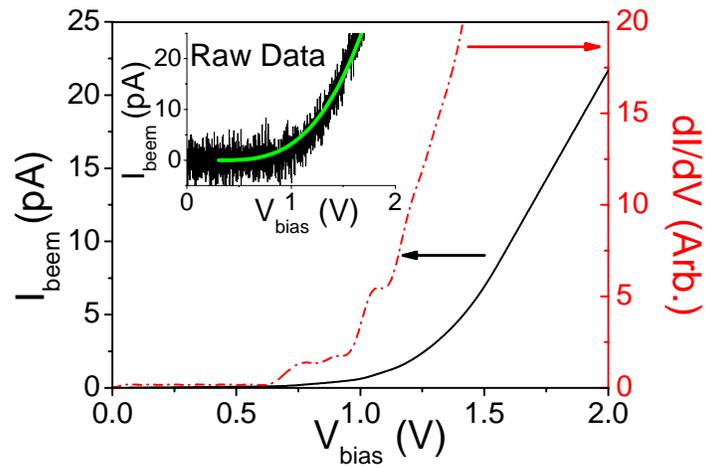

Figure 3. The I-V and the dI/dV of the Ag-PPP interface. Multiple thresholds evidenced as steps in the dI/dV are clearly seen. Inset shows one single spectrum as recorded.

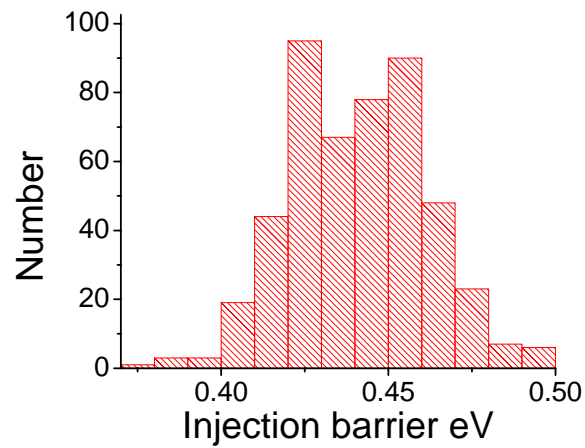

Fig. 4. Distribution of Schottky barrier heights.

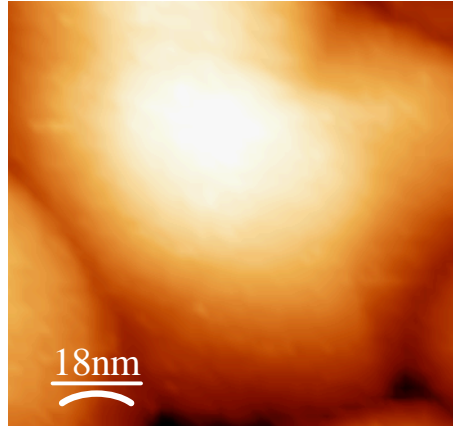

(a)

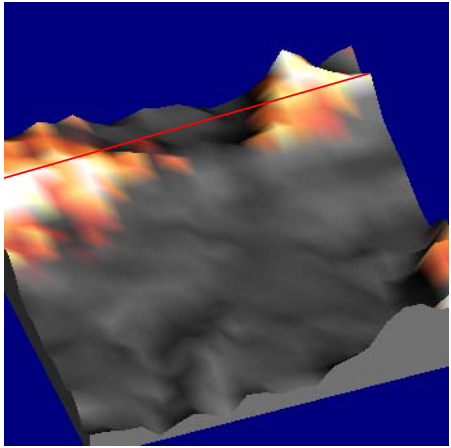

(b)

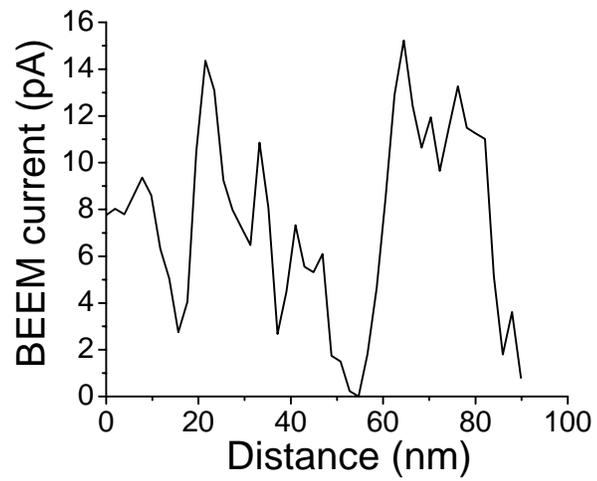

(c)

Figure 5 STM and BEEM current images (a) STM image of a 50 nm square region. The topography scale is 1.2 nm (b) 3D representation of BEEM current from the corresponding region at 1 V bias, with a full scale of 15 pA (c) Profile of BEEM current as a function of position.

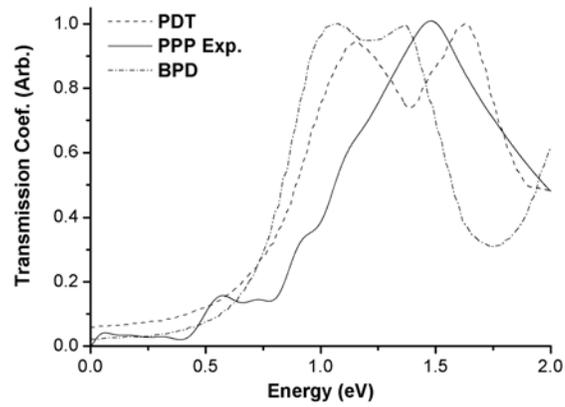

Figure 6. Transmission across a metal-organic interface. Comparison of experimental data, with the calculation by Xue and Ratner [Ref. 24]. Dash dot, dash and solid lines denote transmission coefficient of BPD, PDT and PPP respectively.